\documentclass[conference]{IEEEtran}
\IEEEoverridecommandlockouts
\usepackage{enumitem}
\usepackage{cite}
\usepackage{amsmath,amssymb,amsfonts}
\usepackage{algorithmic}
\usepackage{dsfont}
\usepackage[ruled, vlined]{algorithm2e}
\usepackage{url}
\usepackage{graphicx}
\usepackage{textcomp}
\usepackage{xcolor}
\def\BibTeX{{\rm B\kern-.05em{\sc i\kern-.025em b}\kern-.08em
    T\kern-.1667em\lower.7ex\hbox{E}\kern-.125emX}}

\newcommand{\rrp}{{\mathbb{R}^{p \times p}}} 
\newcommand{\gggg}{\mathcal{G}} 
\newcommand{\vvv}{\mathcal{V}} 
\newcommand{\eee}{\mathcal{E}} 
\newcommand{\cb}{\mathbf{C}} 
\newcommand{\mathbfit}[1]{\textbf{\textit{#1}}} 

\usepackage{todonotes}

\begin{document}

\title{Benchmarking Brain Connectivity Graph Inference: A Novel Validation Approach\\
\thanks{This work was supported by the Agence Nationale de la Recherche under
the France 2030 programme, reference ANR-23-IACL-0006.}}

\author{\IEEEauthorblockN{Alice Chevaux\IEEEauthorrefmark{1}, Ali Fahkar\IEEEauthorrefmark{1}, Kévin Polisano\IEEEauthorrefmark{1}, Irène Gannaz\IEEEauthorrefmark{3}, Sophie Achard\IEEEauthorrefmark{1}} \\
\IEEEauthorblockA{
\IEEEauthorrefmark{1} Univ. Grenoble Alpes, CNRS, Inria, LJK, F-38000 Grenoble, France \\
\IEEEauthorrefmark{3} Univ. Grenoble Alpes, CNRS, Grenoble INP, G-SCOP, 38000 Grenoble, France }
}

\maketitle

\begin{abstract}
Inferring a binary connectivity graph from resting-state fMRI data for a single subject requires making several methodological choices and assumptions that can significantly affect the results. In this study, we investigate the robustness of existing edge detection methods when relaxing a common assumption: the sparsity of the graph. We propose a new pipeline to generate synthetic data and to benchmark the state of the art in graph inference. Simulated correlation matrices are designed to have a set of given zeros and a constraint on the signal-to-noise ratio. We compare approaches based on covariance or precision matrices, emphasizing their implications for connectivity inference. This framework allows us to assess the sensitivity of connectivity estimations and edge detection methods to different parameters.
\end{abstract}


\section{Introduction}
Representing brain connectivity as a graph is highly valuable in neuroscience, as it allows for a structured visualization where nodes correspond to brain regions and edges represent interactions between them. Usually, the brain is parcelled in regions of interest (ROIs) and we exploit fMRI data to provide statistical insights and a clear representation of the dependencies between these brain regions. A key challenge lies in defining the dependencies between brain regions. Since temporal signals are extracted from brain regions, various estimators —such as correlations or partial correlations— can be used to quantify these dependencies. Most methods aim to infer the graph by identifying the zeros of the covariance matrix or the precision matrix. Developing reliable statistical techniques for this inference is essential for ensuring accurate connectivity graph for further analysis. Previous methods have been proposed to evaluate the robustness of connectivity estimations relying essentially on a data driven approach \cite{dadi2019benchmarking, vavsa2022null}, or under the assumption of sparsity \cite{lafit2019partial}.

The aim of this paper is to evaluate different methods for inferring a connectivity graph between brain regions from resting-state fMRI wavelets for a single subject by providing a large set of simulation data. These methods are usually separated in two steps: (i) choosing an estimator of a connectivity measures, (ii) transforming these continuous connectivity measures matrices into an adjacency matrix, where a 1 indicates the presence of a connection, and with 0 otherwise, defining at the same time a connectivity graph. This binarization process, which we refer to as \textit{edge detection methods}, determines which statistical associations should be retained as edges in the graph. This step is often called \textit{sparsification}, but we find this term misleading as it suggests an inherently sparse graph structure, which may not always be the case for fMRI connectivity \cite{markov2014weighted}.

Our study is structured around these two key steps: (i) evaluating the quality of different connectivity indicators and (ii) assessing the sensitivity of edge detection methods. The first part of our results focuses on quantifying how well different indicators (correlation, partial correlation, and their variants) discriminate the zeros and the ones of the ground-truth adjacency matrix. The second part highlights the difficulty to find a robust threshold. Finally, we provide a comparative analysis of several existing calibrated methods for binarizing connectivity matrices with respect to several parameters.

\section{Model Specification}

\subsection{Formalization in a Probabilistic Setting}
In brain connectivity we suppose that the ROIs are associated with $p$ variables that define a random vector $\mathbfit{X}=(X_1,\dots, X_p)^\top$ from which we have $T$  i.i.d realisations that follow a multivariate centered Gaussian distribution, with $\mathbf{\Sigma} \in \rrp$ the matrix of covariance. A matrix $\mathbf{\Sigma}$ from $\rrp$ can be associated to an adjacency matrix defined by $A_{i,j}=\mathbf{1}_{\Sigma_{i,j}\neq 0}$. This allows us to navigate between the space of matrices and the space of graph: a graph $\gggg$ is a mathematical structure used to represent pairwise relations between objects. Formally, a graph is defined as  $\gggg = (\vvv, \eee)$, where $\vvv$ is the set of nodes, and $\eee \subset \vvv \times \vvv$ is the set of edges. Here we want to obtain a connectivity graph, where the nodes are the brain regions and the edges represent the dependencies between them: we want the edges to exist depending on the adjacency matrix from either the covariance matrix $\mathbf{\Sigma}$ or of the precision matrix $\mathbf{\Sigma}^{-1}$. The choice of covariance or precision matrix depends on the kind of dependencies that we want to represent between the ROIs. 


In practice we need to estimate the adjacency matrices of either covariance or precision matrices. We usually prefer to estimate the correlations and the partial correlations. That is $\mathbf{R}=(\rho_{ij})_{i,j=1,\dots,p}=(\text{diag}({\mathbf{ \Sigma}}))^{-\frac{1}{2}} \ {\mathbf{\mathbf{\Sigma}}} \ (\text{diag}({\mathbf{\Sigma}}))^{-\frac{1}{2}}$ for the correlation matrix and its inverse for the partial correlation matrix. We can use the classic empirical estimators: $\hat{\mathbf{\Sigma}}=\frac{\mathbfit{X}^\top \mathbfit{X}}{T}$ and naturally the estimator of the correlations defined by $\hat{\mathbf{R}}=(\hat\rho_{ij})_{i,j=1,\dots,p}=(\text{diag}(\hat{\mathbf{ \Sigma}}))^{-\frac{1}{2}} \ \hat{\mathbf{\mathbf{\Sigma}}} \ (\text{diag}(\hat{\mathbf{\Sigma}}))^{-\frac{1}{2}}$ and its inverse for the empirical partial correlation matrix. These estimators converge towards the correlation and partial correlation matrices as $T$ goes to infinity, however since we are working with a number of observations that can be low, it may be useful to consider other estimators. For instance, the Ledoit-Wolf regularization for empirical covariance is a well-known method that is supposed to reduce the variance of the estimation when the matrix is sparse \cite{ledoit2003honey}.

\subsection{Key Parameters}
For this paper, we want to evaluate the sensibility of the previous connectivity measures and of edge detection methods regarding several parameters identified in previous studies: graph density, signal-to-noise ratio and sample size. We are particularly interested in these parameters because most methods are built on strong assumptions about them.
\begin{itemize}[left=0.2em]
    \item \textbf{Graph density} ($d$): It is often assumed that the true underlying graph is sparse, with the proportion of edges in the adjacency matrix $d$ at most $d=0.05$ \cite{drton2017structure}. We believe that this assumption may be unrealistic for fMRI data \cite{markov2014weighted}.
    
    \item \textbf{Signal-to-noise level} ($b$): Some methods incorporate prior assumptions or tune their parameters based on an expected contrast between zero and nonzero values in the connectivity matrix. They implicitly assume that when an edge is present, its value should be sufficiently distinct from zero. We propose to define the mean value of the nonzero coefficients in the correlation or precision matrix, denoted $b$, as a parameter to control in the simulation of data. This parameter behaves like a signal-to-noise ratio in our study.
    
    \item \textbf{Sample size} ($T$): Some methods are theoretically guaranteed to converge to the true adjacency matrix as the number of observations $T$ increases. In practice, fMRI data are limited samples. We focus on evaluating the performance of the methods when $T \approx 100$, which is a realistic order of magnitude for fMRI wavelet coefficients \cite{bullmore2004wavelets}.
\end{itemize}

\subsection{Synthetic Data Generation}
The difficulty in simulating data for this paper relies on the ability to control the parameter $b$ and $d$ for a positive-definite (PSD) matrix. Using chordal graphs, we are able to simulate graphs $(\vvv,\eee)$ respecting any given proportion of edges $d$ meaning: $d=\frac{|\eee|}{p(p-1)/2}$. For each of these graphs, we want to find a positive-definite matrix respecting its set of edges $\eee$, but we also want to control the mean value of its nonzero coefficients. Convex optimization results from \cite{fakhar2025generating} are used. The optimization problem is defined with two constraints. First, we look for a covariance (resp. precision) matrix $\mathbf{\Sigma}$ associated with a graph $(\vvv,\eee)$, meaning it satisfies the constraint \begin{equation}
	\label{eq:constraint1}
	\mathbf{\Sigma} = (\Sigma_{ij})  \text{ PSD }, \quad \Sigma_{ij} = 0, \ (i, j) \notin \eee.
\end{equation}

We also want to vary the signal-to noise ratio therefore we impose the following constraint:  
\begin{equation}\label{eq:b}\frac{1}{2\vert \eee\vert}\sum_{i\neq j} \Sigma_{ij} \geq b.\end{equation}

We seek to solve the following optimization problem:
\begin{equation} \label{eq:constraint2}
\begin{aligned}
& \underset{\cb}{\text{minimize}} & & \frac{1}{2}\lVert \mathbf{\Sigma} - \bar{\mathbf{\Sigma}} \rVert^2_F, \\
& \text{subject to} & &\text{constraints } \eqref{eq:constraint1} \text{ and } \eqref{eq:b},
\end{aligned}
\end{equation}
with $\bar{\mathbf{\Sigma}}$ a given arbitrary matrix that we want to approach. Since the objective function in~\eqref{eq:constraint2} is convex, a solution exists whenever the constraints are feasible.

\begin{figure}[htbp]
  \begin{minipage}{0.2\textwidth}
    \centerline{\includegraphics[width=3cm,trim={6,7cm 0cm 7cm 0cm},clip]{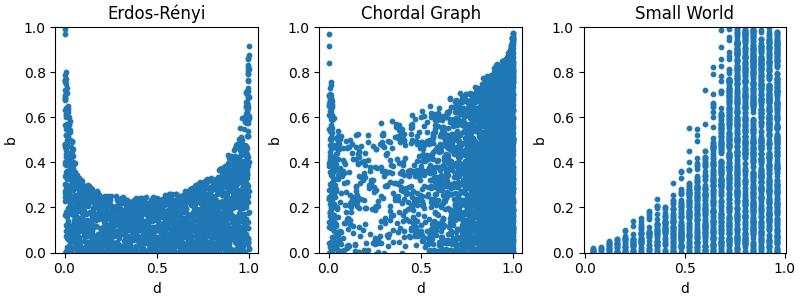}
    }
    
  \end{minipage}
  \begin{minipage}[c]{0.28\textwidth}
    \caption{Representation of the set of matrices we were able to simulate with respect to the mean value of the non-zeros coefficients $b$ and the proportion of edges $d$, with a chordal graph structure.}
    \label{fig:chordalgraph}
  \end{minipage}
  
\end{figure}

To create our datasets we choose to set the dimension of the matrices at $p=51$ (based on a real world fMRI data application that motivates this study). Figure \ref{fig:chordalgraph} gives the values of $b$ and $d$ where we are able to provide matrices satisfying the constraints.

We choose 300 matrices among these represented here, with $b>0.2$. For each generated matrix $\mathbf{\Sigma}$, we can simulate $T$ i.i.d realisations of the centered Gaussian vector $\mathbfit{X}=(X_1,\dots, X_p)^\top$ where the matrix $\mathbf{\Sigma}$ is either the precision or the covariance matrix. We consider here several number of realisations, $T \in \{100,500,1000\}$, to see the robustness of the procedures to the sample size. 

\section{Edges detection methods}
We focus on edge detection models that construct a binary connectivity matrix by estimating correlations or partial correlations based on i.i.d. data. 

\subsection{Hypothesis Testing}
Statistically we can construct an hypothesis test for each edge: for each pair $(i,j)$ $H_0^{i,j}: \rho_{i,j}=0$ and $H_1^{i,j}: \rho_{i,j} > 0$. Here, since we have supposed that the data is Gaussian, we focus on Pearson correlation coefficients. Other tests like permutation tests can be used with less assumptions on the data. We are in a configuration of multiple testing \cite{goeman2014multiple}.
When computing several tests at a risk $\alpha$, we obtain the set $\mathcal{R}=\{(i,j), H_0^{i,j} \text{ is rejected}\}$ using the same data, so we need to apply a correction to control the error at the global level.
\begin{itemize}[left=0.2em]
    \item \textbf{Control of the FWER}: the Family Wise Error Rate (FWER) is defined as: $FWER=\mathbb{P}(\exists (i,j) \in \mathcal{R},\rho_{i,j}=0)$. Representing the probability of having at least one pair that have been considered significantly correlated wrongly. To control this probability at a level $\alpha$, we can apply the Bonferroni procedure \cite{bonferroni1936teoria}.
    \item \textbf{Control of the FDR}: the False Discovery Rate (FDR) is defined as: $FDR=\mathbb{E}[\mathcal{Q}]$ where $\mathcal{Q}=\frac{\{|(i,j)\in \mathcal{R}, \rho_{i,j}=0 |\}}{|\mathcal{R}|\wedge 1}$. This quantifies the expectation of the ratio in $\mathcal{R}$ to have been wrongly rejected. Several procedures exist to control the FDR at a level $\alpha$ like Benjamini-Yekutieli procedure \cite{yekutieli1999resampling}. We note that Benjamini-Hochberg is regularly used for correlation study, however it supposes independence or restrictive dependence structures between the data used for each test, which are {\it a priori} not satisfied in our framework.
\end{itemize}
Multiple testing with correction procedures are known to be very conservative, indeed by construction they favor the null hypothesis. Therefore, in practice, many proposed new methods that are constructed to be less conservative of the null hypothesis, but with no control on the risk.

\subsection{Thresholding}
The most classic way to detect the edges is to apply a threshold on the correlations or partial correlations estimators. The choice of the threshold is either arbitrary or calibrated on the data:

\begin{itemize}[left=0.2em]
\item {\bf Fixed threshold.} An arbitrary threshold between 0 and 1 is applied to the correlation or partial correlation estimator $\hat \rho_{i,j}$. This method is the most commonly used as it is easy to interpret and calculate. It is the main method in the toolbox of neuroscientists such as in \cite{wang2015gretna}. However there is no existing consensus about the choice of the threshold.

\item {\bf Fixed proportion.} A threshold is applied that guarantees the same proportion of edges for each estimated graph. This is used in practice to compare a group of patients with a group of controls. Some articles, such as \cite{van2017proportional}, have warned practitioners about the risks of this method, because the total functional connectivity activity is an information that is erased in this procedure.

\item  {\bf Mixture-Model threshold.} Several methods such as \cite{chen2016bayesian} and \cite{bielczyk2018thresholding} have proposed to model the values of interest (correlations, partial correlations or transforms of those quantities) as mixture models. They suppose that the values come from a mixture of two distributions: one representing the zeros and the other the significant edges. They apply a threshold based on the distribution of zeros they estimate.

\item {\bf Percolation threshold.}  Alternative methods have been developed to find the optimal threshold to apply community-detection method. They focus more on the graph properties to find a threshold. For instance, \cite{bordier2017graph} choose the greater threshold that, once applied to the Pearson correlation matrix, still provides a connected graph.
\end{itemize}

\subsection{Sparse Gaussian Graphical Model}

To avoid finding the optimal threshold, we can directly use a method that estimates a covariance or precision matrix with zeros. The most common one is based on Gaussian graphical models. Using these models, zeros of the matrices are directly estimated. However, it is needed to have a Gaussian assumption on the data which is not the case for multiple testing which require only an hypothesis on asymptotic normality.
\begin{itemize}[left=0.2em]
\item  {\bf Graphical Lasso.} This method provides a sparse estimator for the precision matrix. Because of the necessity to calibrate a regularization parameter using cross-validation, this method may be sensitive to small sample sizes $T$. 
\end{itemize}
Other graphical Gaussian models exist, but they do not give better results when directly estimating a matrix with zeros as shown in \cite{lafit2019partial}. Finally, a similar method has been developed to estimate the covariance matrix in \cite{bien2011sparse}, but without applications to fMRI and with a time consuming algorithm.

\section{Results}
\subsection{Metrics of Performance }
The strength of this paper relies on our simulation process, that allows us to have for each dataset a ground-truth graph $\mathcal{G}^*$ respecting the set of parameters $(d,b,T)$. It defines a ground-truth adjacency matrix $\mathbf{A}^*$: each edge $(i,j)$ is either present ($A^*_{ij} = 1$) or absent ($A^*_{ij} = 0$). During most edge detection methods, we first estimate connectivity indicators $\hat{\cb}$, where each edge $(i,j)$ has a measure $c_{ij}$. We can compare these connectivity indicators using the Area Under the Curve.

\textbf{Area Under the Curve (AUC)}: the AUC evaluates how well the connectivity indicators $\{c_{ij}\}$ rank the true edges higher than the non-edges. AUC = 1 indicates perfect ranking, AUC = 0.5 corresponds to random ranking. It is defined as the probability that a randomly chosen true edge $(i,j)$ has a higher score than a randomly chosen non-edge $(k,l)$:
\[
\text{AUC} = \mathbb{P}(c_{ij} > c_{kl} \mid A^*_{ij} = 1, A^*_{kl} = 0) \approx
 \frac{R_1 - \frac{N_1 (N_1 + 1)}{2}}{N_1 N_0},
\]
where $N_1$ and $N_0$ are the numbers of positive and negative instances, respectively, and $R_1$ is the sum of the ranks of the positive instances when sorting all scores in ascending order.

Once an edge detection is applied to obtain a binary estimated graph $\hat{\mathcal{G}}$ with adjacency matrix $\hat{\mathbf{A}}$, we define the following indicators of quality:
\begin{itemize}[left=0.2em]
    \item \textbf{Accuracy}: $\displaystyle \frac{TP + TN}{TP + TN + FP + FN}$
    \item \textbf{True Positive Rate (TPR)}: $\displaystyle \frac{TP}{TP + FN}$
    \item \textbf{False Positive Rate (FPR)}: $\displaystyle \frac{FP}{FP + TN}$
\end{itemize}
where $TP$ (True Positives) are correctly detected edges ($\hat{A}_{ij} = 1, A^*_{ij} = 1$), $TN$ (True Negatives) are correctly absent edges ($\hat{A}_{ij} = 0, A^*_{ij} = 0$), $FP$ (False Positives) are incorrectly added edges ($\hat{A}_{ij} = 1, A^*_{ij} = 0$), and $FN$ (False Negatives) are missed edges ($\hat{A}_{ij} = 0, A^*_{ij} = 1$).

\subsection{Comparison of the Connectivity Estimators}
\begin{figure}[htbp]
\centerline{\includegraphics[width=9cm]{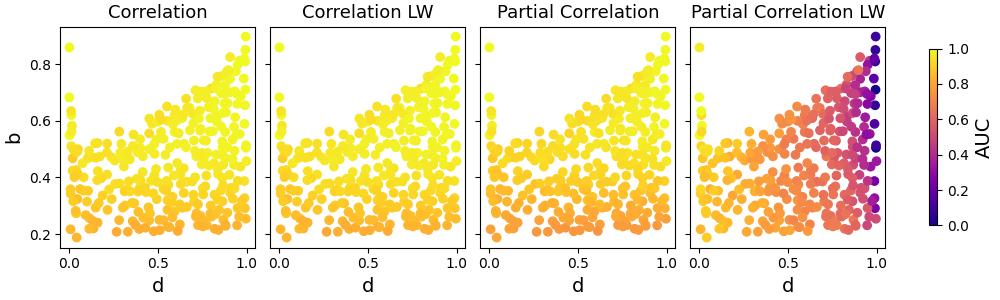}}
\caption{Comparison of the AUC of different connectivity indicators when $T=100$, depending on the mean of the non-zero coefficients ($b$) and the graph density ($d$).}
\label{fig:auc}
\end{figure}

A first evaluation of the differences between the estimators used for either correlations or partial correlations is illustrated in Figure \ref{fig:auc} using the AUC. We have seen that we can estimate them with common empirical estimators or using Ledoit-Wolf regularization. A high AUC means that the measure is able to greatly recover the adjacency matrix by thresholding, subject to finding the optimum threshold, which is the focus of the next part.

In Figure \ref{fig:auc}, we observe that empirical partial correlations and empirical correlations yield similar results, both being influenced by the value of $b$, as expected. Still, empirical partial correlations perform slightly worse. The Ledoit-Wolf correlation matrix estimator does not provide a clear improvement, even for small values of $d$, meaning the ranking of connectivity measures remains unchanged. When using Ledoit-Wolf to estimate partial correlations, we lose a significant amount of information, especially for large $d$, likely due to the smoothing effect applied to the covariance matrix.

\subsection{Impact of the Parameters on the Threshold}

Now that we have an idea of the performance of the connectivity measures to discriminate the zeros, we want to represent directly the accuracy obtained when applying different thresholds between 0 and 1. Indeed, even if the connectivity indicators are highly discriminative, it does not help us to find an optimal threshold. Here we present only the study with the empirical correlations, but the results where globally the same with partial correlations.  
\begin{figure}[htbp]
\centerline{\includegraphics[width=9cm]{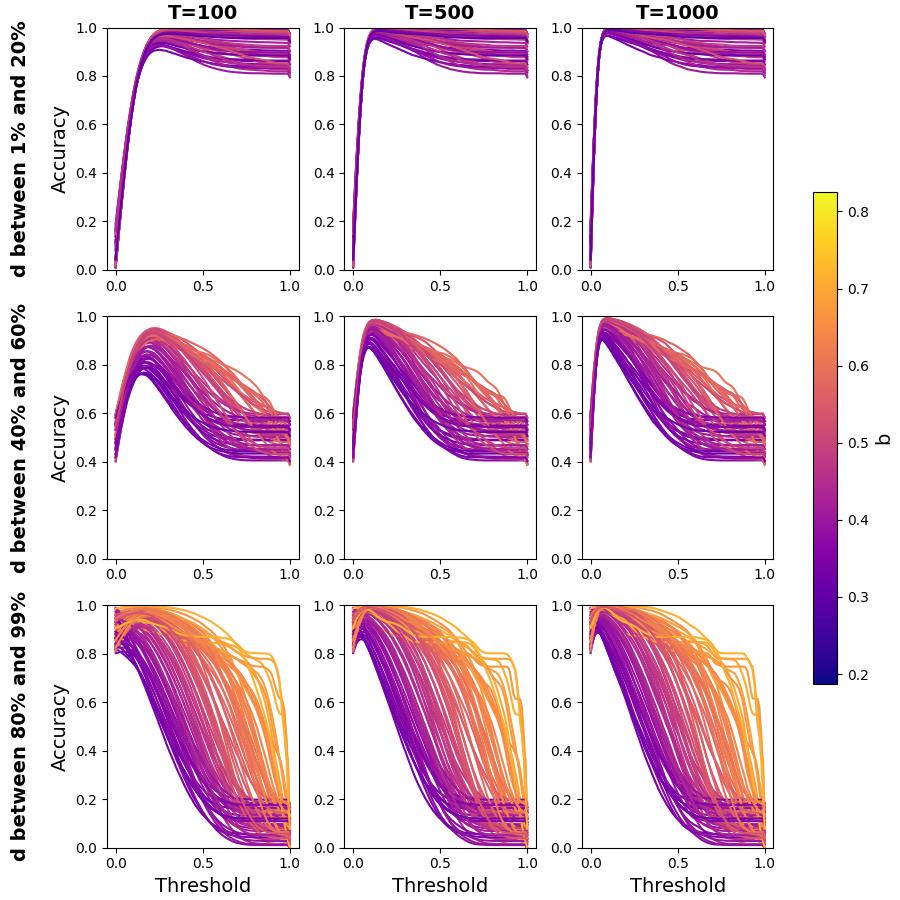}}
\caption{Accuracy obtained when applying different thresholds on an empirical correlation matrix, depending on the graph density $d$, the number of observations $T$, and the mean value of the non-zeros coefficients $b$.}
\label{fig:thresholdcurve}
\end{figure}

Figure \ref{fig:thresholdcurve} displays the accuracy with respect to the applied threshold. 
 The objective is to show how the optimal threshold for accuracy changes, depending on the parameters $d$, $b$ and $T$. When the threshold is near 0, the estimated graph is full, thus the accuracy is equal to $d$, the proportion of existing edges. When the threshold is near 1, the accuracy is equal to $1-d$, the proportion of zeros in the ground-truth graph. The peak of accuracy arises when we indeed detect meaningful edges. To make the effects of $d$ clear, we represent the curves obtained for the matrices for different range of $d$ (depending on the row). Meanwhile the columns depend on the number of realizations (from $T=100$ to $T=1000$). Finally the color of the curve is associated with the value of $b$.

First, the value of $d$ impacts the choice of the optimal threshold as we can see that the differences of shape for these curves depend on $d$. We can also see that when $T$ increases, the optimal threshold is closer to zero. This is directly explained by the fact that the empirical correlation converges correctly to zero if there is no edge. Finally, the higher the value of $b$, the wider the range of thresholds where the accuracy is good. These results emphasize how thin the range of useful thresholds is, and how the three parameters considered ($d$, $b$ and $T$) should be taken into account when choosing a threshold.

\subsection{Comparison of the Edge Detection Methods}
The three edge detection methods compared here are the multiple testing approach with Bonferroni procedure, the thresholding procedure with a Percolation threshold, and the Graphical Lasso with a cross-validation step. This way we have one method for each types of edge detection methods. We choose to represent once again the 300 matrices by their value of $b$ and $d$, this way we can represent by color different measures of performance for the edges detection methods: Accuracy, False Positive Rate and True Positive Rate.

\begin{figure}[htbp]
\centerline{\includegraphics[width=9cm]{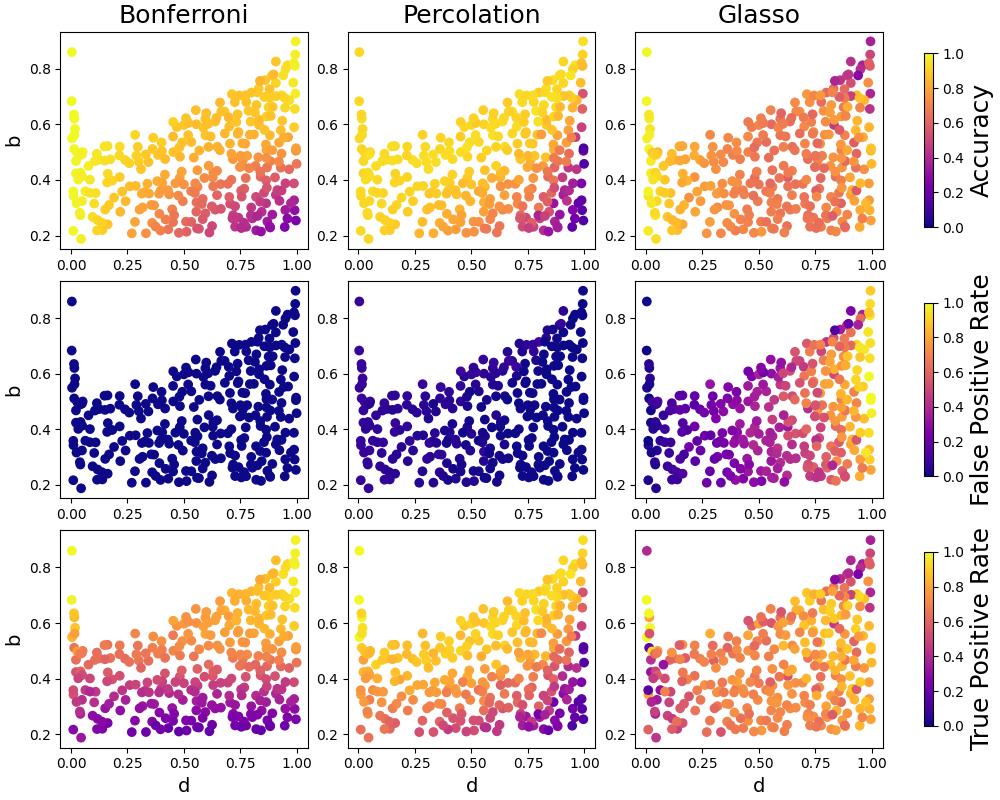}}
\caption{Comparison of the Bonferonni procedure, the Graphical Lasso and the Percolation threshold methods (from left to right) when $T=100$. The y-axis is the mean value of non-zeros coefficients ($b$) and the x-axis is the graph density ($d$). Several metrics are considered (from top to bottom): Accuracy, False Positive Rate (FPR) and True positive Rate (TPR).}
\label{accuracytprfdr100}
\end{figure}

These are the notable behaviour of the methods based on the Figure \ref{accuracytprfdr100} when $T=100$:
\begin{itemize}[left=0.2em]
\item Graphical Lasso: If the graph is not sparse, this method struggles to provide accurate results when $T=100$.

\item Percolation threshold: Its performance appears to depend more on $d$ than on $b$. This is expected, as the method is designed to focus solely on the graph structure.

\item Multiple testing with Bonferonni correction: Considering accuracy alone, one might conclude that the Percolation threshold method and multiple testing behave similarly. However, a closer look at the true positive rate (TPR) reveals a key difference: it is sensitive to $d$ for the Percolation threshold method, while it is solely influenced by $b$ for the multiple testing approach.
\end{itemize}

\begin{figure}[htbp]
\centerline{\includegraphics[width=9cm]{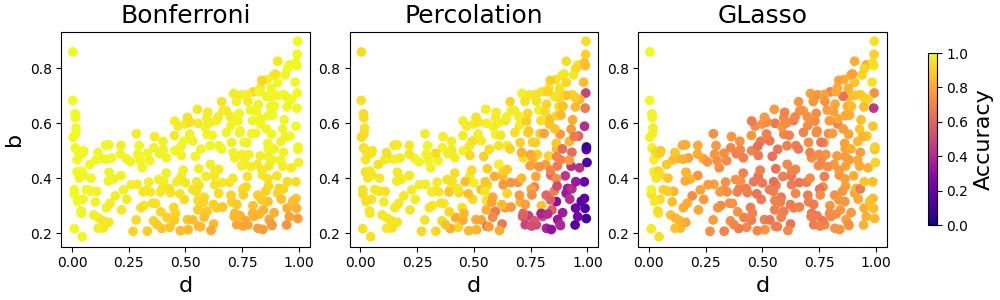}}
\caption{Comparison of the Bonferonni procedure, the Graphical Lasso and the Percolation threshold methods (from left to right) when T=1000: we plot the mean value of non-zeros coefficients (d) depending on the degree of density (d) and color by the accuracy.}
\label{accuracytprfdr1000}
\end{figure}

Figure \ref{accuracytprfdr1000} further demonstrates that these methods behave differently as $T$ increases since we have now $T=1000$:
\begin{itemize}[left=0.2em]
\item Bonferroni multiple testing and Graphical Lasso follow the same pattern as before for Accuracy but with improved overall performance.

\item Percolation thresholding does not significantly improve results. While it removes noise-induced edges, it also eliminates true edges for large $d$ since it prioritizes sparsity while preserving connectivity. Unlike other methods, Percolation thresholding is based on the graph structure rather than on statistical properties of the correlation estimator. Therefore, the issue persists even for high $T$. 
\end{itemize}

\section{Conclusion and Perspectives}

Our benchmark provides a simple, efficient and original method to evaluate the accuracy of the binarization of connectivity matrices with ground truth. Codes and datasets are available to reproduce the experiments\footnote{\url{https://gricad-gitlab.univ-grenoble-alpes.fr/polisank/benchmarking-brain-connectivity-graph-inference-a-novel-validation-approach/}}. No sparsity assumptions are needed and this covers different graph configurations, including those observed in neuroimaging. Our analysis highlights that the optimal threshold is not a universal constant but rather depends on at least three key parameters: $d$, $b$, and $T$. Any chosen method must take these effects into account to ensure reliable and interpretable results. As future work, we need to extend to other graph structures including small-world and preferential attachment for example. Finally, weighted graphs may be also used in the context of neuroimaging. Weights are also affected by errors in the estimations, and the same strategy of validation can be used with methods inferring weighted graphs.

\bibliographystyle{IEEEtran}
\bibliography{references.bib}

\end{document}